\documentclass[twocolumn,english,showpacs,preprintnumbers,prx]{revtex4-1}
\usepackage[latin9]{inputenc}
\usepackage{color}
\usepackage{amsmath}
\usepackage{amssymb}
\usepackage{graphicx}
\usepackage{esint}

\makeatletter

\@ifundefined{textcolor}{}
{%
 \definecolor{BLACK}{gray}{0}
 \definecolor{WHITE}{gray}{1}
 \definecolor{RED}{rgb}{1,0,0}
 \definecolor{GREEN}{rgb}{0,1,0}
 \definecolor{BLUE}{rgb}{0,0,1}
 \definecolor{CYAN}{cmyk}{1,0,0,0}
 \definecolor{MAGENTA}{cmyk}{0,1,0,0}
 \definecolor{YELLOW}{cmyk}{0,0,1,0}
}

\@ifundefined{textcolor}{}{%
 \definecolor{BLACK}{gray}{0}
 \definecolor{WHITE}{gray}{1}
 \definecolor{RED}{rgb}{1,0,0}
 \definecolor{GREEN}{rgb}{0,1,0}
 \definecolor{BLUE}{rgb}{0,0,1}
 \definecolor{CYAN}{cmyk}{1,0,0,0}
 \definecolor{MAGENTA}{cmyk}{0,1,0,0}
 \definecolor{YELLOW}{cmyk}{0,0,1,0}
}

\@ifundefined{textcolor}{}{%
 \definecolor{BLACK}{gray}{0}
 \definecolor{WHITE}{gray}{1}
 \definecolor{RED}{rgb}{1,0,0}
 \definecolor{GREEN}{rgb}{0,1,0}
 \definecolor{BLUE}{rgb}{0,0,1}
 \definecolor{CYAN}{cmyk}{1,0,0,0}
 \definecolor{MAGENTA}{cmyk}{0,1,0,0}
 \definecolor{YELLOW}{cmyk}{0,0,1,0}
}

\@ifundefined{textcolor}{}{%
 \definecolor{BLACK}{gray}{0}
 \definecolor{WHITE}{gray}{1}
 \definecolor{RED}{rgb}{1,0,0}
 \definecolor{GREEN}{rgb}{0,1,0}
 \definecolor{BLUE}{rgb}{0,0,1}
 \definecolor{CYAN}{cmyk}{1,0,0,0}
 \definecolor{MAGENTA}{cmyk}{0,1,0,0}
 \definecolor{YELLOW}{cmyk}{0,0,1,0}
}

\@ifundefined{textcolor}{}{%
 \definecolor{BLACK}{gray}{0}
 \definecolor{WHITE}{gray}{1}
 \definecolor{RED}{rgb}{1,0,0}
 \definecolor{GREEN}{rgb}{0,1,0}
 \definecolor{BLUE}{rgb}{0,0,1}
 \definecolor{CYAN}{cmyk}{1,0,0,0}
 \definecolor{MAGENTA}{cmyk}{0,1,0,0}
 \definecolor{YELLOW}{cmyk}{0,0,1,0}
}

\@ifundefined{textcolor}{}{%
 \definecolor{BLACK}{gray}{0}
 \definecolor{WHITE}{gray}{1}
 \definecolor{RED}{rgb}{1,0,0}
 \definecolor{GREEN}{rgb}{0,1,0}
 \definecolor{BLUE}{rgb}{0,0,1}
 \definecolor{CYAN}{cmyk}{1,0,0,0}
 \definecolor{MAGENTA}{cmyk}{0,1,0,0}
 \definecolor{YELLOW}{cmyk}{0,0,1,0}
}

\@ifundefined{textcolor}{}{%
 \definecolor{BLACK}{gray}{0}
 \definecolor{WHITE}{gray}{1}
 \definecolor{RED}{rgb}{1,0,0}
 \definecolor{GREEN}{rgb}{0,1,0}
 \definecolor{BLUE}{rgb}{0,0,1}
 \definecolor{CYAN}{cmyk}{1,0,0,0}
 \definecolor{MAGENTA}{cmyk}{0,1,0,0}
 \definecolor{YELLOW}{cmyk}{0,0,1,0}
}

\@ifundefined{textcolor}{}{%
 \definecolor{BLACK}{gray}{0}
 \definecolor{WHITE}{gray}{1}
 \definecolor{RED}{rgb}{1,0,0}
 \definecolor{GREEN}{rgb}{0,1,0}
 \definecolor{BLUE}{rgb}{0,0,1}
 \definecolor{CYAN}{cmyk}{1,0,0,0}
 \definecolor{MAGENTA}{cmyk}{0,1,0,0}
 \definecolor{YELLOW}{cmyk}{0,0,1,0}
}

\usepackage{ulem}

\usepackage{aecompl}

\usepackage{epsfig}\usepackage{dcolumn}\usepackage{bm}

\usepackage{babel}

\usepackage{babel}

\makeatother

\usepackage{babel}
\begin{document}

\title{Catching the bound states in the continuum of a phantom atom in graphene}

\author{L. H. Guessi$^{1}$, R. S. Machado$^{2}$, Y. Marques$^{2}$, L.
S. Ricco$^{2}$, K. Kristinsson$^{3},$ M. Yoshida$^{1}$, I. A. Shelykh$^{3,4,5}$,
M. de Souza$^{1,},$}

\altaffiliation{Current address: Institute of Semiconductor and Solid State Physics, Johannes Kepler University Linz, Austria.}

\author{A. C. Seridonio$^{1,2}$}

\affiliation{$^{1}$IGCE, Unesp - Univ Estadual Paulista, Departamento de F\'{i}sica,
13506-900, Rio Claro, SP, Brazil\\
 $^{2}$Departamento de F\'{i}sica e Qu\'{i}mica, Unesp - Univ Estadual
Paulista, 15385-000, Ilha Solteira, SP, Brazil\\
 $^{3}$Division of Physics and Applied Physics, Nanyang Technological
University 637371, Singapore\\
 $^{4}$Science Institute, University of Iceland, Dunhagi-3, IS-107,
Reykjavik, Iceland\\
 $^{5}$ ITMO University, St. Petersburg 197101, Russia}
\begin{abstract}
We explore theoretically the formation of bound states in the continuum
(BICs) in graphene hosting two collinear adatoms situated at different
sides of the sheet and at the center of the hexagonal cell, where
a phantom atom of a fictitious lattice emulates the six carbons of
the cell. We verify that in this configuration the local density of
states (LDOS) near the Dirac points exhibits two characteristic features:
i) the cubic dependence on energy instead of the linear one for graphene
as found in New J. Phys. \textbf{16}, 013045 (2014) and ii) formation
of BICs as aftermath of a Fano destructive interference assisted by
the Coulomb correlations in the adatoms. For the geometry where adatoms
are collinear to carbon atoms, we report absence of BICs.
\end{abstract}


\maketitle

\section{Introduction}

\label{sec:SecI}

Graphene is a two-dimensional material consisting of an atomic monolayer
where carbon atoms build a honeycomb lattice, which is characterized
by a band structure exhibiting a massless relativistic dispersion
relation in the vicinity of the Dirac cones situated at the corners
of the Brillouin zone \cite{Novo1,Peres,Neto}. Recent experimental
and theoretical works demonstrated the possibility of the effective
controllable adsorption of impurities, the so-called adatoms, by an
individual graphene sheet \cite{Eelbo1,Eelbo2,abInitio1}. These astonishing
hallmarks have driven researchers towards a topic of the electron
tunneling through adatoms in a relativistic environment \cite{Uchoa1,Uchoa2,Uchoa3}.
The variety of the adatom geometries considered so far and novel effects
predicted are quite broad. For instance, in a system composed by a
couple of magnetic adatoms, the exchange coupling results in a highly
anisotropic RKKY interaction \cite{RKKY3,RKKY4}.

In this context, the Scanning Tunneling Microscope (STM) technique
has been recognized as the most efficient experimental tool \cite{STMreview}.
Its use allows to probe the local density of state (LDOS) of the system.
Interestingly enough, the latter is governed by the Fano interference
effect \cite{Fano2} between the direct tunneling from the STM tip
to the host and that via the adatom. In addition, the Fano effect
forms the basis of the appearance of the so-called bound states in
the continuum (BICs).

\begin{figure}[!]
\includegraphics[width=0.5\textwidth,height=0.2\textheight]{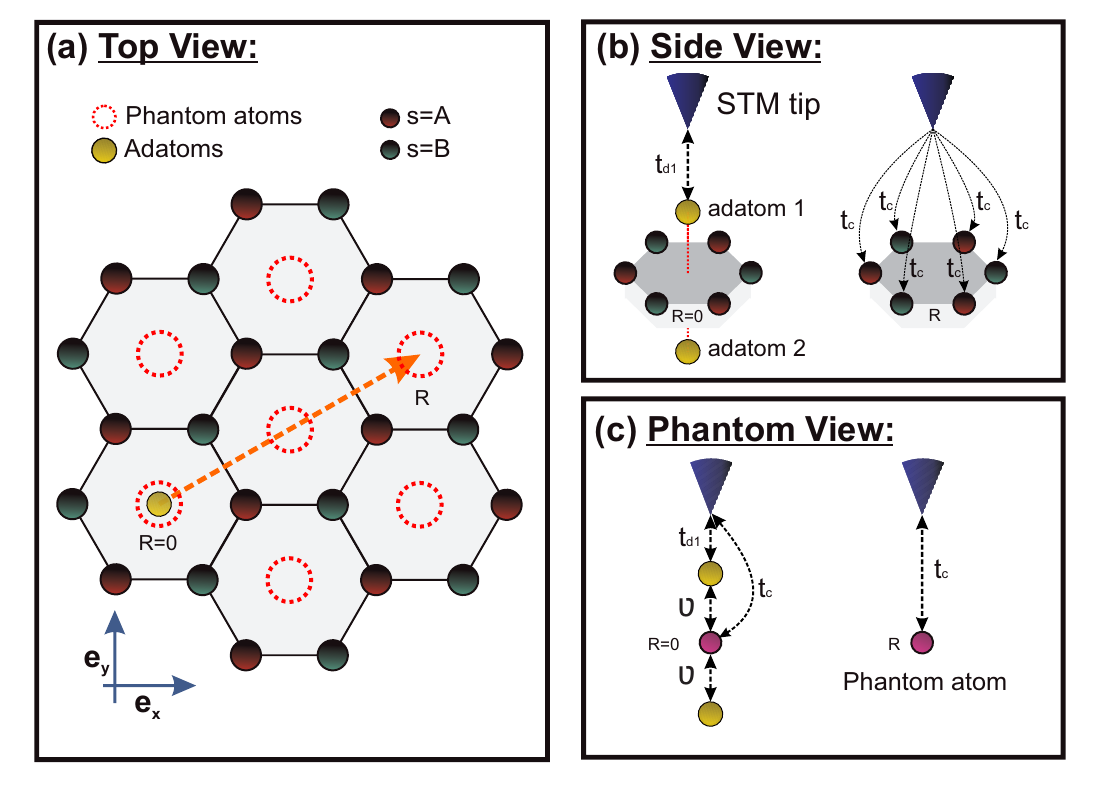}
\protect\protect\protect\protect\protect\protect\protect\protect\protect\protect\protect\protect\protect\protect\protect\protect\protect\protect\protect\protect\protect\protect\protect\caption{\label{fig:Pic1} (Color online) The geometry of the system we consider.
(a) The dotted-red circles represent a fictitious lattice composed
by phantom atoms in graphene. (b) At the position ${\bold R}={\bold0}$
the STM tip couples to the adatom 1 and the six atoms of carbon (only
shown for an arbitrary ${\bold R}$). (c) Phantom atoms (shaded-red
spheres) emulating the cells of (b)\textcolor{blue}{.}}
\end{figure}

BICs were first theoretically predicted by von Neumann and Wigner in
1929 \cite{Neuman-1} as quantum states with localized square-integrable
wave functions appearing above the threshold of a given stationary
potential. The solutions of the corresponding Schrödinger equation
are characterized by destructive interference between partial waves
which cancel the amplitude of the wave function at large distances
from the potential core. Notably, the subject received a revival after
the publication of the work of Stillinger and Herrick in 1975 \cite{SH}.
Since then, appearance of BICs was predicted in optical and photonic
systems \cite{Exp1,Boretz,Exp2,Exp3}, setups with peculiar chirality
\cite{Chiral}, Floquet-Hubbard states induced by a strong oscillating
electric field \cite{Floquet} and driven by A.C. fields \cite{AC},
among others.

In the domain of the carbon-based structures, graphene ribbons were
proposed as appropriate candidates for the detection of BICs \cite{Gonzalez,Gong}.
However, from the perspective of quantum transport, such states are
difficult to see. Indeed, as the electrons within BICs are not allowed
to leak into the continuum, they become invisible in transport experiments.
Hence, in order to proof the existence of BICs, proposals of novel
experimental setups suitable for their detection are of fundamental
interest.

In this article we discuss theoretically the necessary conditions
for the appearance of BICs in graphene-adatom systems. We show that
such states appear if two collinear adatoms with Coulomb correlations
are placed above and below the center of the hexagonal cell as shown
at Fig.\ref{fig:Pic1}. The situation can be considered by means of
the introduction of a fictitious, or phantom atom located at the center
of the hexagonal cell and coupled to the STM tip in the transport
experiment. In this configuration, the formation of the BIC is assisted
by a Fano interference mechanism. Similar process takes place in the
optical and photonic systems described in Refs.\,{[}\onlinecite{Boretz}{]}
and {[}\onlinecite{Exp3}{]}. The phantom atom belongs to a fictitious
lattice composed by atoms of the same species with DOS presenting a cubic
energy dependency as it was originally predicted by B. Uchoa \textit{et
al}\textit{\textcolor{blue}{.}}\textcolor{blue}{{} }\cite{Uchoa2,Uchoa3}\textcolor{blue}{.}
To make the BIC visible, one needs to introduce the mechanism of its
coupling with the continuum, which can be done by the use of a detuning
between the energy levels of the adatoms.

\section{The model}

\label{sec:SecII}

To give a theoretical description of the setup presented at Fig.\ref{fig:Pic1},
we develop the model based on the two-impurity Anderson Hamiltonian
treated in frameworks of Hubbard I approximation \cite{Hubbard}.
The system is described by the model Hamiltonian
\begin{equation}
\mathcal{H}_{\text{{T}}}=\mathcal{H}_{\text{{2D}}}+\mathcal{H}_{\text{tip}}+\mathcal{H}_{\text{tun}}.
\end{equation}
The first term of $\mathcal{H}_{\text{{T}}}$ represents the Anderson
like-model:
\begin{align}
\mathcal{H}_{\text{{2D}}} & =\mathcal{H}_{\text{{g}}}+\mathcal{H}_{\text{{d}}}+\mathcal{H}_{\mathcal{V}},\label{eq:TIAM}
\end{align}
where the first part corresponds to the free graphene sheet
\begin{equation}
\mathcal{H}_{\text{{g}}}=-t\sum_{\langle\bar{m},m\rangle\sigma}[\hat{\Psi}_{A\sigma}^{\dagger}({\bold R}_{\bar{m}})\hat{\Psi}_{B\sigma}({\bold R}_{m})+\text{{H.c.}}]
\end{equation}
in which $\langle\bar{m},m\rangle$ runs over the nearest neighbors
of carbon atoms with hopping term $t\approx2.8$\,eV, $\hat{\Psi}_{s\sigma}^{\dagger}({\bold R}_{m})$
($\hat{\Psi}_{s\sigma}({\bold R}_{m})$) is the creation (annihilation)
fermionic operator of an electron for a given spin $\sigma$ in a
sublattice $s=A,B.$
\begin{equation}
\mathcal{H}_{\text{{d}}}=\sum_{j\sigma}\mathcal{E}_{jd\sigma}n_{d_{j}\sigma}+\mathcal{U}\sum_{j}n_{d_{j}\uparrow}n_{d_{j}\downarrow}
\end{equation}
describes the adatoms $(j=1,2)$, where $n_{d_{j}\sigma}=d_{j\sigma}^{\dagger}d_{j\sigma}$,
$d_{j\sigma}^{\dagger}$ ($d_{j\sigma}$) creates (annihilates) an
electron with spin $\sigma$ in the state $\mathcal{E}_{jd\sigma}=\mathcal{E}_{d}+(-1)^{1-j}\Delta\mathcal{E}$
with the index $j=1,2$ designating the upper and lower adatoms respectively,
$\Delta\mathcal{E}$ represents the possible detuning between the
levels of the different adatoms and $\mathcal{U}$ accounts for the
on-site Coulomb interaction.
\begin{equation}
\mathcal{H}_{\mathcal{V}}=\mathcal{V}\sum_{j=1}^{2}\sum_{i=1}^{3}\sum_{\sigma}\{[\hat{\Psi}_{A\sigma}({\bold\delta_{i}})+\hat{\Psi}_{B\sigma}(-{\bold\delta_{i}})]d_{j\sigma}^{\dagger}+\text{{H.c.}}\}\label{eq:HV}
\end{equation}
hybridizes the six atoms of the hexagonal cell with the couple of
adatoms as sketched in Fig. \ref{fig:Pic1}. ${\bold\delta_{1}}=a{\bold e}_{x}$
and ${\bold\delta}_{2,3}=\frac{a}{2}(-{\bold e}_{x}\pm\sqrt{3}{\bold e}_{y})$
represent the nearest neighbor vectors of carbon atoms, $a\sim1.4$\,${\AA}$
is the distance between graphene atoms and $\mathcal{V}$ is the hybridization
strength\textcolor{blue}{, }which is supposed to be the same for the
six carbons of the hexagonal cell. This assumption holds for adatoms
with orbital symmetry $s,$ $f_{z^{3}}$ and $d_{z^{2}}$ (Co atoms
for instance) \cite{Uchoa3}.

The second part of $\mathcal{H}_{\text{{T}}}$ is described by the
Hamiltonian $\mathcal{H}_{\text{tip}},$ which corresponds to free
electrons in the STM tip. The tunneling Hamiltonian, describing the
tip-host coupling can be expressed as
\begin{eqnarray}
\mathcal{H}_{\text{tun}}=\sum_{\sigma}[t_{\text{{c}}}\Psi_{\sigma}({\bold R})+t_{\text{{d1}}}d_{1\sigma}]\Psi_{\text{tip}\sigma}^{\dagger}+\text{{H.c.}}=\nonumber \\
=t_{\text{{c}}}\sum_{\sigma}\tilde{\Psi}_{\sigma}({\bold R})\Psi_{\text{tip}\sigma}^{\dagger}+\text{{H.c.},}\label{eq:HtunII}
\end{eqnarray}
where $\Psi_{\text{tip}\sigma}$ is the operator for the edge site
of the tip and
\begin{align}
\Psi_{\sigma}({\bold R}) & =\sum_{i=1}^{3}[\hat{\Psi}_{A\sigma}(\bold R+{\bold\delta_{i}})+\hat{\Psi}_{B\sigma}(\bold R-{\bold\delta_{i}})]\label{eq:phantom_1}
\end{align}
describes the six carbon atoms of the hexagonal cell with its center
collinear to the STM tip position ${\bold R}$ as outlined at Fig.\ref{fig:Pic1}.
The field operator
\begin{equation}
\tilde{\Psi}_{\sigma}({\bold R})=\Psi_{\sigma}({\bold R})+(t_{\text{{d1}}}/t_{\text{{c}}})d_{1\sigma}\label{eq:FieldO}
\end{equation}
accounts for the quantum interference between the direct electron
tunneling through the carbons of such a cell and tunneling through
the adatom $1$ placed above the central site of the cell.\textcolor{blue}{{}
}Note that for the ratio $t_{\text{{d1}}}/t_{\text{{c}}}\rightarrow0$
the coupling of the adatom $1$ to the STM is negligible compared to
the tip-host coupling. The achievement of this regime can be reliable
by the employment of an atom with deeply localized orbital. Such an
orbital is characterized by a wave function which is more compact
than that of carbon atoms, thus preventing that the hopping term $t_{\text{{d1}}}$
becomes dominant.

After some algebra \cite{diagonal}, Eq.(\ref{eq:phantom_1}) can
be reduced to
\begin{align}
\Psi_{\sigma}({\bold R=\bold0}) & =\frac{1}{2\pi}\sqrt{\frac{\pi\Omega_{0}}{\mathcal{N}}}\sum_{ns}\int\left(\frac{\hbar v_{F}k}{-t}\right)\sqrt{\left|k\right|}dkc_{nsk\sigma}\nonumber \\
 & \equiv\Psi_{\text{{phantom},}\sigma},\label{eq:phantom_2}
\end{align}
which corresponds to the fermionic operator describing the quantum
state of the fictitious or phantom atom placed in the center of the
hexagonal cell, where $n$ runs over the Dirac points $\bold K_{\pm}=2\pi/3a(1,\pm1/\sqrt{3}).$

By applying the linear response theory, in which the STM tip is considered
as a probe, it is possible to show that the differential conductance
is determined by
\begin{equation}
G({\bold R})\sim\frac{e^{2}}{h}\pi\Gamma_{\text{{tip}}}\text{{LDOS}}({\bold R}),\label{eq:Conduc}
\end{equation}
where $e$ is the electron charge, $\Gamma_{\text{{tip}}}=4\pi t_{c}^{2}\rho_{\text{tip}}$,
$\rho_{\text{tip}}$ is the DOS for the tip and $\text{{LDOS}}({\bold R})$
is the LDOS of the phantom atom perturbed by the adatoms, which despite
being a local property it accounts for the entire bath composed by
the phantom atoms. It is worth mentioning that if one increases the
ratio $t_{\text{{d1}}}/t_{\text{{c}}}$ in Eq.(\ref{eq:FieldO}),
one should treat the coupling to STM at the same footing as the coupling
in ``graphene+adatoms'' system and as a result, the conductance is not simply
proportional to the LDOS as predicted by the linear
response theory (Eq.(\ref{eq:Conduc})). For the regime of strong
coupling between adatom and STM tip, the theoretical framework found
in Ref.\,{[}\onlinecite{Uchoa3}{]} can be applied for the calculation
of the conductance. However, as we do not expect the appearance of
the BICs in this situation, its detailed analysis is outside the scope
of the current work.

To obtain such a LDOS we first change the system Hamiltonian of Eq.(\ref{eq:TIAM})
to the momenta domain by performing the transformation
\begin{equation}
\Psi_{s\sigma}({\bold R}_{m})=\frac{1}{\sqrt{\mathcal{N}}}\sum_{\bold k}e^{i{\bold k}\cdot{\bold R}_{m}}c_{s\bold k\sigma},
\end{equation}
with $\mathcal{N}$ as the total number of states, $c_{A\bold k\sigma}=a_{\bold k\sigma}$
and $c_{B\bold k\sigma}=b_{\bold k\sigma}$, which yields the Hamiltonian:
\begin{eqnarray}
\mathcal{H}_{\text{{2D}}}=-t\sum_{\bold k\sigma}[\phi({\bold k})a_{\bold k\sigma}^{\dagger}b_{\bold k\sigma}+\text{H.c.}]+\sum_{j\sigma}\mathcal{E}_{jd\sigma}n_{d_{j}\sigma}\nonumber \\
+\mathcal{U}\sum_{j}n_{d_{j}\uparrow}n_{d_{j}\downarrow}+\mathcal{V}\sum_{j\sigma}[\Psi_{\sigma}({\bold R}={\bold0})d_{j\sigma}^{\dagger}+\text{H.c.}],\nonumber \\
\label{eq:H2DII}
\end{eqnarray}
where
\begin{eqnarray}
\Psi_{\sigma}({\bold R})=\frac{1}{\sqrt{\mathcal{N}}}\sum_{\bold k}e^{i\bold k.\bold R}(\phi({\bold k})a_{\bold k\sigma}+\phi^{*}({\bold k})b_{\bold k\sigma})\nonumber \\
\label{eq:PsiSigma}
\end{eqnarray}
and $\phi({\bold k})=\sum_{i=1}^{3}e^{i{\bold k}\cdot{\bold\delta}_{i}}.$

Next we introduce the retarded Green's function in time domain $\tau$
\begin{equation}
\mathcal{G}_{\sigma}({\bold R},\tau)=-\frac{i}{\hbar}\theta\left(\tau\right){\tt Tr}\{\varrho_{\text{2D}}[\tilde{\Psi}_{\sigma}({\bold R},\tau),\tilde{\Psi}_{\sigma}^{\dagger}({\bold R},0)]_{+}\},
\end{equation}
where $\theta\left(\tau\right)$ is the Heaviside function, $\varrho_{\text{2D}}$
is the density matrix of the system described by the Hamiltonian of
Eq.(\ref{eq:TIAM}) and $[\cdots,\cdots]_{+}$ is the anticommutator
between operators taken in the Heisenberg picture.

Therefore, the LDOS can be obtained as
\begin{equation}
\text{{LDOS}}({\bold R})=-\frac{1}{\pi}{\tt Im}[\sum_{\sigma}\tilde{\mathcal{G}}_{\sigma}({\bold R},\mathcal{E}^{+})],
\end{equation}
where $\tilde{\mathcal{G}}_{\sigma}({\bold R},\mathcal{E}^{+})$ is
the time Fourier transform of $\mathcal{G}_{\sigma}({\bold R},\tau).$
Then by applying the equation of motion (EOM) to the $\mathcal{G}_{\sigma}({\bold R},\tau),$
one can show that near the Dirac points where $t|\phi({\bold k})|=\hbar v_{F}k$
one has:
\begin{eqnarray}
\text{{LDOS}}({\bold R})=2\text{\ensuremath{\mathcal{D}}}_{0}+\Delta\text{{LDOS}}({\bold R})=\label{eq:LDOSp0}\\
=2\text{\ensuremath{\mathcal{D}}}_{0}+\sum_{jl}\Delta\text{{LDOS}}_{jl}({\bold R}).\nonumber
\end{eqnarray}
Here
\begin{equation}
\text{\ensuremath{\mathcal{D}}}_{0}\equiv\text{\ensuremath{\mathcal{D}}}_{0}^{\text{{phantom}}}=\frac{1}{\mathcal{N}}\frac{\Omega_{0}}{\pi(\hbar v_{F})^{2}}\frac{|\mathcal{E}|^{3}}{t^{2}}\label{eq:phantom_DOS}
\end{equation}
corresponds to the DOS of the fictitious lattice of so-called phantom
atoms as depicted in Fig.\ref{fig:Pic1}(a). It is worth noticing
that such a DOS is spatially independent as expected for a translational
invariant system, thus revealing that the aforementioned lattice is
periodic over a set of phantom atoms and encloses all energy continuum.
This DOS is expressed in terms of the Fermi velocity $v_{F}$ and
the unit cell area $\Omega_{0}.$

The induced density of states reads
\begin{align}
\Delta\text{{LDOS}}_{jl}({\bold R}) & =-\Delta\text{\ensuremath{\mathcal{D}}}_{0}\sum_{\sigma}{\tt Im}\{[q_{j}({\bold R})-i\mathcal{F}_{j}({\bold R})]\tilde{\mathcal{G}}_{d_{l\sigma}d_{j\sigma}}\nonumber \\
 & \times[q_{l}(-{\bold R})-i\mathcal{F}_{l}(-{\bold R})]\}.\label{eq:LDOSp1}
\end{align}
It is coordinate dependent, which is a clear consequence of the breaking
of the periodicity of the phantom lattice due to the presence
of the adatoms. Clearly, it depends on the Green's functions of the
adatoms, namely $\tilde{\mathcal{G}}_{d_{l\sigma}d_{j\sigma}}$ ($j,l=1,2$),
which can be obtained by determining the time Fourier transform of

{}
\begin{equation}
\mathcal{G}_{d_{l\sigma}d_{j\sigma}}(\tau)=-\frac{i}{\hbar}\theta\left(\tau\right){\tt Tr}\{\varrho_{\text{2D}}[d_{l\sigma}(\tau),d_{j\sigma}^{\dagger}(0)]_{+}\}.\label{eq:Gjl}
\end{equation}
Eq.(\ref{eq:LDOSp1}) also depends on the position ${\bold R}$ of
the phantom atom, the Anderson broadening{} $\Delta=\pi\text{\ensuremath{\mathcal{D}}}_{0}^{\text{{phantom}}}\mathcal{V}^{2}\propto|\mathcal{E}|^{3},$
which according to Ref.{[}\onlinecite{Uchoa3}{]} arises from adatoms
with electronic orbitals obeying the $C_{3v}$ group symmetry as for
instance the cases $s,$ $f_{z^{3}}$ and $d_{z^{2}}.$
\begin{equation}
q_{j}({\bold R})=\frac{1}{\Delta}\text{{\tt Re}}\Sigma_{\text{{phantom}}}({\bold R})+\delta_{j1}(\pi\Delta\text{\ensuremath{\mathcal{D}}}_{0})^{-1/2}(t_{\text{{d1}}}/t_{\text{{c}}})\label{qj}
\end{equation}
is the Fano factor that characterizes the interference between the
direct adatom-host and STM-host paths \cite{Fano2} defined by the
ratio $t_{\text{{d1}}}/t_{\text{{c}}}$. The factor $\mathcal{F}_{j}({\bold R})$
reads:
\begin{equation}
\mathcal{F}_{j}({\bold R})=-\frac{1}{\Delta}\text{{\tt Im}}\Sigma_{\text{{phantom}}}({\bold R}),
\end{equation}
where \textcolor{blue}{{}}
\begin{equation}
\Sigma_{\text{{phantom}}}({\bold R})=\frac{2\mathcal{V}^{2}}{\mathcal{N}}\sum_{\bold k}\frac{e^{-i\bold k.\bold R}\mathcal{E}^{+}|\phi({\bold k})|^{2}}{\mathcal{E}^{+2}-t^{2}|\phi({\bold k})|^{2}}\label{Sigma}
\end{equation}
is the self-energy, which at ${\bold R=\bold0}$ and near the Dirac
points can be approximated by {}
\begin{equation}
\Sigma_{\text{{phantom}}}({\bold R=\bold0})=2\mathcal{V}^{2}\frac{\mathcal{E}}{D^{2}t^{2}}(\mathcal{E}^{2}\ln\Big|\frac{\mathcal{E}^{2}}{D^{2}-\mathcal{E}^{2}}\Big|-D^{2})-i\Delta\label{Sigma0}
\end{equation}
as it was originally derived in Refs.\,{[}\onlinecite{Uchoa2}{]}
and {[}\onlinecite{Uchoa3}{]} ($D\approx7$\,eV denotes the band-edge).

From the point of view of the STM-host coupling, a phantom atom emulates
a single site beneath the STM tip. Note that the $\text{\ensuremath{\mathcal{D}}}_{0}$
of Eq.(\ref{eq:phantom_DOS}) differs from the standard DOS of graphene
in the situation of a single carbon connected to a tip, which is characterized
by $\text{\ensuremath{\mathcal{D}}}_{0}^{\text{{carbon}}}=\Omega_{0}|\mathcal{E}|/2\mathcal{N}\pi(\hbar v_{F})^{2}.$
The cubic dependence $\sim|\mathcal{E}|^{3}$ at low energies for
the phantom DOS arises from the quantum interference between the electron
paths through the hexagonal cell: the straight aftermath of such a
process is the modification of the band-structure of graphene, thus
distorting the well-known linear behavior for the DOS when the STM
tip position coincides with the center of the hexagon.

To determine the density of states ${DOS}_{jj}$ of the adatoms at the site
${\bold R=\bold0}$ of the host we should calculate the Green's functions
$\tilde{\mathcal{G}}_{d_{j\sigma}d_{j\sigma}}$:
\begin{equation}
\text{DOS}_{jj}=-\frac{1}{\pi}{\tt Im}(\sum_{\sigma}\tilde{\mathcal{G}}_{d_{j\sigma}d_{j\sigma}}).\label{DOS}
\end{equation}
To this end, the Hubbard I approximation can be used \cite{Hubbard}.
This approach provides reliable results away from the Kondo regime
\cite{Kondo3}. We start by employing the equation-of-motion (EOM)
method to a single particle retarded Green's function of Eq.(\ref{eq:Gjl})
in time domain for an adatom. Going to energy domain one gets: {}
\begin{align}
(\mathcal{E}^{+}-\mathcal{E}{}_{ld\sigma})\tilde{\mathcal{G}}_{d_{l\sigma}d_{j\sigma}} & =\delta_{lj}+\Sigma_{\text{{phantom}}}({\bold R=\bold0})\sum_{\tilde{l}}\tilde{\mathcal{G}}_{d_{\tilde{l}\sigma}d_{j\sigma}}\nonumber \\
 & +\mathcal{U}\tilde{\mathcal{G}}_{d_{l\sigma}n_{d_{l}\bar{\sigma}},d_{j\sigma},}\label{eq:s1}
\end{align}
with $\mathcal{E}^{+}=\mathcal{E}+i0^{+}.$ In the equation above,
$\tilde{\mathcal{G}}_{d_{l\sigma}n_{d_{l}\bar{\sigma}},d_{j\sigma}}$
denotes a two particle Green's function composed by four fermionic
operators, obtained by the Fourier transform of
\begin{equation}
\mathcal{G}_{d_{l\sigma}n_{d_{l}\bar{\sigma}},d_{j\sigma}}=-\frac{i}{\hbar}\theta\left(\tau\right){\tt Tr}\{\varrho_{\text{2D}}[d_{l\sigma}\left(\tau\right)n_{d_{l}\bar{\sigma}}\left(\tau\right),d_{j\sigma}^{\dagger}\left(0\right)]_{+}\},
\end{equation}
where $\bar{\sigma}=-\sigma$ and $n_{d_{l}\bar{\sigma}}=d_{l\bar{\sigma}}^{\dagger}d_{l\bar{\sigma}}$.
In order to close the system of the dynamic equations, we obtain the
EOM for the Green's function $\mathcal{G}_{d_{l\sigma}n_{d_{l}\bar{\sigma}},d_{j\sigma}}$,
which reads:
\begin{align}
(\mathcal{E}^{+}-\mathcal{E}_{ld\sigma}-\mathcal{U})\tilde{\mathcal{G}}_{d_{l\sigma}n_{d_{l}\bar{\sigma}},d_{j\sigma}} & =\delta_{lj}<n_{d_{l}\bar{\sigma}}>\nonumber \\
+\sum_{\bold ks}\frac{\mathcal{V}}{\sqrt{\mathcal{N}}}[-\phi_{s}({\bold k})\tilde{\mathcal{G}}_{c_{s\bold k\bar{\sigma}}^{\dagger}d_{l\bar{\sigma}}d_{l\sigma},d_{j\sigma}} & +\phi_{s}^{*}({\bold k})(\tilde{\mathcal{G}}_{c_{s\bold k\sigma}d_{l\bar{\sigma}}^{\dagger}d_{l\bar{\sigma}},d_{j\sigma}}\nonumber \\
+\tilde{\mathcal{G}}_{d_{l\bar{\sigma}}^{\dagger}c_{s\bold k\bar{\sigma}}d_{l\sigma},d_{j\sigma}}) & ],\label{eq:H_GF_2}
\end{align}
where the index $s=A,B$ marks a sublattice, $c_{A\bold k\sigma}=a_{\bold k\sigma}$
and $c_{B\bold k\sigma}=b_{\bold k\sigma}$, $\phi_{A}({\bold k})=\phi^{*}({\bold k})$
and $\phi_{B}({\bold k})=\phi({\bold k})$, expressed in terms of
new Green's functions of the same order of $\tilde{\mathcal{G}}_{d_{l\sigma}n_{d_{l}\bar{\sigma}},d_{j\sigma}}$
and the occupation number can be determined as
\begin{equation}
<n_{d_{l}\bar{\sigma}}>=-\frac{1}{\pi}\int_{-D}^{+D}n_{F}(\mathcal{E}){\tt Im}(\tilde{\mathcal{G}}_{d_{l{\bar{\sigma}}}d_{l{\bar{\sigma}}}})d\mathcal{E},
\end{equation}
with $n_{F}(\mathcal{E})$ as the Fermi-Dirac distribution.

Our approach holds for temperatures $T\gg T_{K}$ (above the Kondo
temperature). However, the temperature should not be very high in
order that we can safely employ the Heaviside step function for the
Fermi-Dirac distribution $n_{F}(\mathcal{E})$ \cite{Uchoa1}. By
employing the Hubbard I approximation, we decouple the Green's functions
in the right-hand side of Eq.(\ref{eq:H_GF_2}), as follows: $\tilde{\mathcal{G}}_{c_{s\bold k\bar{\sigma}}^{\dagger}d_{l\bar{\sigma}}d_{l\sigma},d_{j\sigma}}\simeq<c_{s\bold k\bar{\sigma}}^{\dagger}d_{l\bar{\sigma}}>\tilde{\mathcal{G}}_{d_{l\sigma}d_{j\sigma}}$
and $\tilde{\mathcal{G}}_{d_{l\bar{\sigma}}^{\dagger}c_{s\bold k\bar{\sigma}}d_{l\sigma},d_{j\sigma}}\simeq<c_{s\bold k\bar{\sigma}}^{\dagger}d_{l\bar{\sigma}}>\tilde{\mathcal{G}}_{d_{l\sigma}d_{j\sigma}}$,
where we have used $\sum_{\bold ks}\phi({\bold k})=\sum_{\bold ks}\phi^{*}({\bold k})$.
As a result, we find
\begin{eqnarray}
(\mathcal{E}^{+}-\mathcal{E}_{ld\sigma}-\mathcal{U})\tilde{\mathcal{G}}_{d_{l\sigma}n_{d_{l}\bar{\sigma}},d_{j\sigma}} & = & \delta_{lj}<n_{d_{l}\bar{\sigma}}>\nonumber \\
+\frac{\mathcal{V}_{j}}{\sqrt{\mathcal{N}}}\sum_{\bold ks} & \phi_{s}^{*}({\bold k}) & \tilde{\mathcal{G}}_{c_{s\bold k\sigma}d_{l\bar{\sigma}}^{\dagger}d_{l\bar{\sigma}},d_{j\sigma}}.\nonumber \\
\label{eq:H_GF_3-1}
\end{eqnarray}
To complete the calculation, we need to determine $\tilde{\mathcal{G}}_{c_{s\bold k\sigma}d_{l\bar{\sigma}}^{\dagger}d_{l\bar{\sigma}},d_{j\sigma}}$.
Once again, employing the EOM approach for $\tilde{\mathcal{G}}_{c_{s\bold k\sigma}d_{l\bar{\sigma}}^{\dagger}d_{l\bar{\sigma}},d_{j\sigma}}$,
we obtain
\begin{align}
\mathcal{E}^{+}\tilde{\mathcal{G}}_{c_{s\bold k\sigma}d_{l\bar{\sigma}}^{\dagger}d_{l\bar{\sigma}},d_{j\sigma}} & =-t\phi_{\bar{s}}({\bold k})\tilde{\mathcal{G}}_{c_{\bar{s}\bold k\sigma}d_{l\bar{\sigma}}^{\dagger}d_{l\bar{\sigma}},d_{j\sigma}}\nonumber \\
+ & \sum_{\bold q\tilde{s}}\frac{\mathcal{V}_{l}}{\sqrt{\mathcal{N}}}\phi_{\tilde{s}}^{*}({\bold q})\tilde{\mathcal{G}}_{c_{s\bold k\sigma}d_{l\bar{\sigma}}^{\dagger}c_{\tilde{s}\bold q\bar{\sigma}},d_{j\sigma}}\nonumber \\
+ & \sum_{\tilde{j}}\frac{\mathcal{V}_{\tilde{j}}}{\sqrt{\mathcal{N}}}\phi_{s}({\bold k})\tilde{\mathcal{G}}_{d_{\tilde{j}\sigma}n_{d_{l}\bar{\sigma}},d_{j\sigma}}\nonumber \\
- & \sum_{\bold q\tilde{s}}\frac{\mathcal{V}_{l}}{\sqrt{\mathcal{N}}}\phi_{\tilde{s}}({\bold q})\tilde{\mathcal{G}}_{c_{\tilde{s}\bold q\bar{\sigma}}^{\dagger}d_{l\bar{\sigma}}c_{s\bold k\sigma},d_{j\sigma}},\nonumber \\
\label{eq:H_GF_4}
\end{align}
where $\bar{s}=A,B$ respectively for $s=B,A$ as labels to correlate
simultaneously distinct sublattices, while $\tilde{s}=A,B$ runs arbitrarily.

In a similar way by using Hubbard I scheme for Eq.(\ref{eq:H_GF_4})
we have $\tilde{\mathcal{G}}_{c_{s\bold k\sigma}d_{l\bar{\sigma}}^{\dagger}c_{\tilde{s}\bold q\bar{\sigma}},d_{j\sigma}}\simeq\left\langle d_{l\bar{\sigma}}^{\dagger}c_{\tilde{s}\bold q\bar{\sigma}}\right\rangle \tilde{\mathcal{G}}_{c_{s\bold k\sigma}d_{j\sigma}},$
$\tilde{\mathcal{G}}_{c_{\tilde{s}\bold q\bar{\sigma}}^{\dagger}d_{l\bar{\sigma}}c_{s\bold k\sigma},d_{j\sigma}}\simeq\left\langle d_{l\bar{\sigma}}^{\dagger}c_{\tilde{s}\bold q\bar{\sigma}}\right\rangle \tilde{\mathcal{G}}_{c_{s\bold k\sigma}d_{j\sigma}}$
and $\tilde{\mathcal{G}}_{d_{\tilde{j}\sigma}n_{d_{l}\bar{\sigma}},d_{j\sigma}}\simeq\left\langle n_{d_{l}\bar{\sigma}}\right\rangle \tilde{\mathcal{G}}_{d_{\tilde{j}\sigma}d_{j\sigma}}$,
which in combination with Eqs.\,(\ref{eq:s1}) and (\ref{eq:H_GF_3-1})
results in

\begin{equation}
\tilde{\mathcal{G}}_{d_{j\sigma}d_{j\sigma}}=\frac{\lambda_{j}^{\bar{\sigma}}}{\mathcal{E}-\mathcal{E}_{jd\sigma}-{{\tilde{\Sigma}}^{\sigma}}_{jj}},\label{eq:pass3-1}
\end{equation}
where $\lambda_{j}^{\bar{\sigma}}=(1+\frac{\mathcal{U}<n_{d_{j}\bar{\sigma}}>}{\mathcal{E}-\mathcal{E}_{jd\sigma}-\mathcal{U}-\Sigma_{\text{{phantom}}}({\bold R=\bold0})})$,
and

{}
\begin{equation}
{{\tilde{\Sigma}}^{\sigma}}_{jj}=\Sigma({\bold R=\bold0})+\lambda_{j}^{\bar{\sigma}}\lambda_{\bar{j}}^{\bar{\sigma}}\frac{[\Sigma_{\text{{phantom}}}({\bold R=\bold0})]^{2}}{\mathcal{E}-\mathcal{E}_{\bar{j}d\sigma}-\Sigma_{\text{{phantom}}}({\bold R=\bold0}))}\label{eq:TSE}
\end{equation}
is the total self-energy, with $\bar{j}=2,1$ respectively for $j=1,2$
for the indexes corresponding to distinct adatoms and
\begin{eqnarray}
\tilde{\mathcal{G}}_{d_{j\sigma}d_{\bar{j}\sigma}} & = & \frac{\lambda_{j}^{\bar{\sigma}}\Sigma_{\text{{phantom}}}({\bold R=\bold0})\tilde{\mathcal{G}}_{d_{\bar{j}\sigma}d_{\bar{j}\sigma}}}{\mathcal{E}-\mathcal{E}_{jd\sigma}-\Sigma_{\text{{phantom}}}({\bold R=\bold0})}\label{eq:G12}
\end{eqnarray}
are mixed Green's functions, which describe the correlations between
the adatoms and are responsible for Fano destructive interference.

\section{Results and Discussion}

\label{sec:SecIII}

In the discussion below we adopt the following set of the system parameters:
$t_{\text{{d1}}}/t_{\text{{c}}}=10^{-6},$ which ensures the assumption
of the STM tip acting as a probe of the ``graphene+adatoms'' system
LDOS as discussed in Sec.\ref{sec:SecII}\textcolor{blue}{, }$\mathcal{E}_{d}=-0.07D,$
$\mathcal{U}=0.14D,$ $\mathcal{V}=0.14D$ and $v_{F}\approx c/300$
\cite{Uchoa1}.

\begin{figure}
\centering{}\includegraphics[width=0.5\textwidth,height=0.45\textheight]{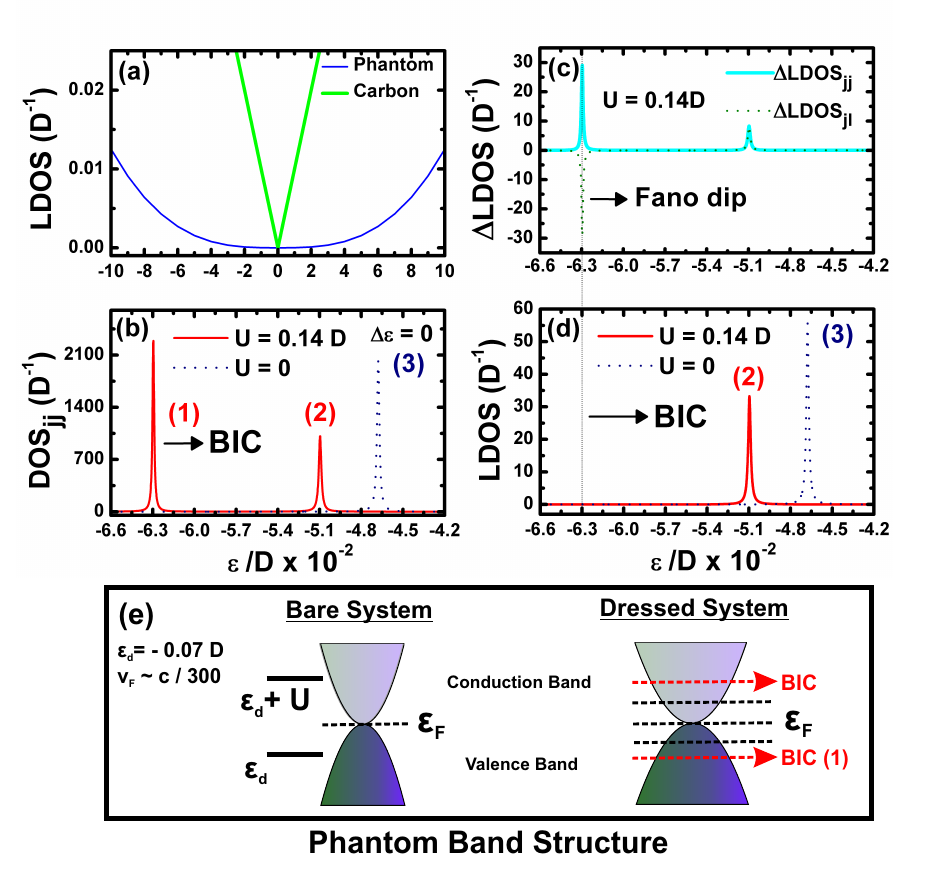}
\protect\protect\protect\protect\protect\protect\protect\protect\protect\protect\protect\protect\protect\caption{\label{fig:Pic2} (Color online) (a) LDOS of graphene coupled to STM
tip for two configurations: the tip above the carbon atom of graphene
(labeled as carbon) and the tip in the center of the hexagonal lattice
(labeled as phantom). (b) Density of states for the pair of adatoms
$\text{DOS}_{jj}=\text{DOS}_{11}=\text{DOS}_{22}$ within the valence
band. The parameters are $\mathcal{E}_{d}=-0.07D$, $\mathcal{U}=0.14D,$
$\mathcal{V}=0.14D,$ $v_{F}\approx c/300$ and $\triangle\mathcal{E}=0.$
Additional two peaks in the conduction band are symmetrically placed
if the condition $2\mathcal{E}_{d}+\mathcal{U}=0$ is satisfied, which
are not shown. (c) Contributions to the LDOS of graphene from the
adatom pair. Diagonal contribution shows two pronounced peaks, while
mixing term shows a single pronounced antiresonance. (d) Total LDOS
revealing the BIC (marked by vertical line) at position where the
resonance of the diagonal term in the LDOS is compensated by the antiresonance
in the mixing term. (e) Sketch of the energy diagram of the system.
Left: energy levels without the dressing of the adatoms by conducting
electrons. Right: the energy diagram accounting for the Coulomb dressing.
The pair of BICs is indicated by red arrows. In both of right and
left panels the coupling of the graphene sheet to the STM tip turns
the linear $|\mathcal{E}|$ dependence within the density of states
of the host into cubic one ($|\mathcal{E}|^{3}$). }
\end{figure}

Panel (a) of the Fig.\ref{fig:Pic2} shows the comparison between
the linear LDOS of graphene (green curve) versus $\text{\ensuremath{\mathcal{D}}}_{0}^{\text{{phantom}}}$
with cubic dependence characteristic for the phantom atom of Eq.(\ref{eq:phantom_DOS})
(blue curve). Fig.\ref{fig:Pic2}(b) displays the densities of states
of the adatoms $\text{DOS}_{jj}=\text{DOS}_{11}=\text{DOS}_{22}$
defined by Eq.(\ref{DOS}) with zero detuning $(\triangle\mathcal{E}=0),$
where two peaks labeled as (1) and (2) are situated within the valence
band $(\mathcal{E}<\mathcal{E}_{F}\equiv0)$ for the case of $\mathcal{U}\neq0$
(red curve). Two extra peaks appear within the conduction band $(\mathcal{E}>\mathcal{E}_{F}\equiv0$)
as well (not shown), since we assumed the symmetric Anderson model
with the constraint $2\mathcal{E}_{d}+\mathcal{U}=0$ being fulfilled.
In this regime, the graphene Hamiltonian with adatoms is invariant
under particle-hole transformation, and all the properties of the
peaks within the conduction band are the same as those within the
valence band, thus we do not need to perform separate analysis for
them. Note, that deviations from the condition $2\mathcal{E}_{d}+\mathcal{U}=0$
will not change the presented results qualitatively, but positions
of the peaks in conduction and valence bands will not be anymore symmetric.
For comparison we also present the curve for $\mathcal{U}=0$ (dark-dotted
curve), characterized by a single peak labeled as (3).

Panel (c) of the Fig.\ref{fig:Pic2} shows the contributions of the
adatoms to the LDOS of graphene. For $\mathcal{U}\neq0$ the diagonal
term $\Delta\text{LDOS}_{jj}=\Delta\text{LDOS}_{11}=\Delta\text{LDOS}_{22}$
displays pronounced peaks at the same energies as the DOS of the adatoms
shown at the panel (b). On the contrary, the mixing term $\Delta\text{LDOS}_{jl}=\Delta\text{LDOS}_{12}=\Delta\text{LDOS}_{21}$
exhibits sharp Fano dip corresponding to the peak located around $\mathcal{E}\approx-6.3\times10^{-2}D$.
When all contributions to the LDOS are added, this antiresonance cancels
exactly the corresponding resonance in the diagonal term. This means
that the peak (1) of panel (b) can be considered as a BIC arising
from a Fano destructive interference assisted by Coulomb correlations:
in the situation of finite Coulomb potential $\mathcal{U},$ $\Delta\text{LDOS}_{jl}$
for $j\neq l$ describes electronic waves that travel forth and back
between the upper and lower adatoms, which for a given energy $\mathcal{E},$
become phase shifted by $\pi$ with respect to the waves scattered
by the adatoms enclosed by $\Delta\text{LDOS}_{jj}.$ Particularly
at the sites of the adatoms where the BICs lie and with $\mathcal{E}\approx-6.3\times10^{-2}D,$
such a condition is fulfilled and is reflected by the peak and Fano
dip, respectively in $\Delta\text{LDOS}_{11}=\Delta\text{LDOS}_{22}$
and $\Delta\text{LDOS}_{12}=\Delta\text{LDOS}_{21}$ as found in Fig.\ref{fig:Pic2}(c).

\begin{figure}
\centering{}\includegraphics[width=0.5\textwidth,height=0.3\textheight]{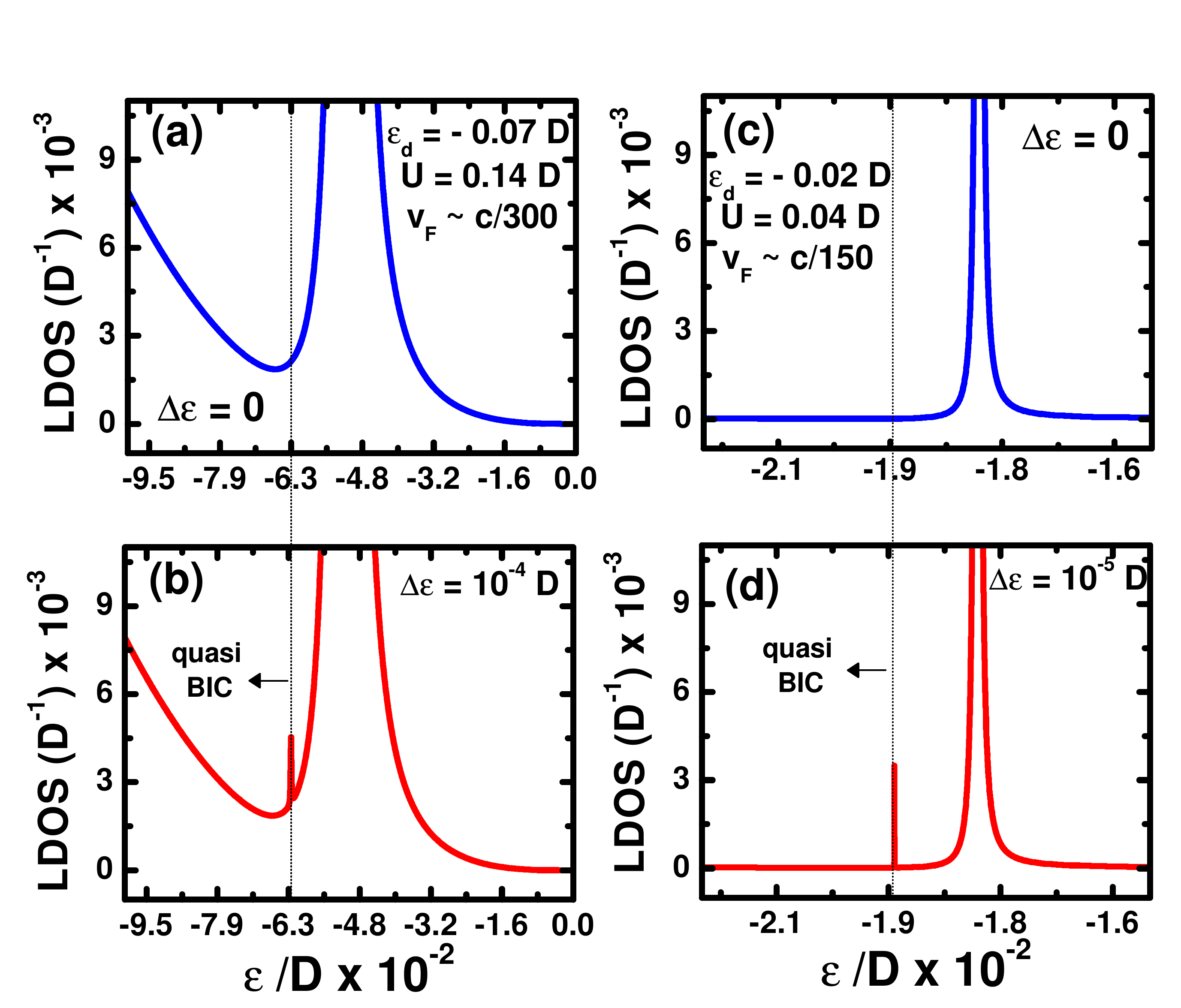}
\protect\protect\protect\protect\protect\protect\protect\protect\protect\protect\protect\protect\protect\caption{\label{fig:Pic3} (Color online) (a) $\mathcal{E}_{d}=-0.07D$, $\mathcal{U}=0.14D,$
$\mathcal{V}=0.14D,$ $v_{F}\approx c/300$ and $\triangle\mathcal{E}=0:$
LDOS in the region around the position of the BIC. The latter lies
at $\mathcal{E}\approx-6.3\times10^{-2}D$ and is invisible in the
LDOS, its position is shown by a vertical line. (b) The LDOS in the
region around the position of the BIC for non-zero detuning between
the energies of the upper and lower adatoms $\triangle\mathcal{E}=10^{-4}D,$
all other parameters are the same as in panel (a). One clearly sees
that the BIC is reflected in the LDOS in form of a tiny peak at $\mathcal{E}\approx-6.3\times10^{-2}D$
and thus should become detectable in transport measurements. (c) Same
as panel (a), but for the different values of parameters: $\mathcal{E}_{d}=-0.02D$,
$\mathcal{U}=0.04D$ and $v_{F}\approx c/150.$ (d) The LDOS in the
region around the position of the BIC for non-zero detuning between
the energies of the upper and lower adatoms $\triangle\mathcal{E}=10^{-5}D,$
all other parameters are the same as in panel (c).}
\end{figure}

We highlight that the peak in the DOS marked as peak (1) in Fig. \ref{fig:Pic2}(b)
appearing around $\mathcal{E}\approx-6.3\times10^{-2}D$ in the red
curve does not rise at the same position in the LDOS of panel (d)
due to the Fano suppression mechanism, thus preventing the revealing
of the BIC by a conductance measurement. This feature is made explicit
by the vertical line crossing both panels (c) and (d) of Fig.\ref{fig:Pic2},
where the BIC position is marked. We checked that for $\mathcal{U}=0$
BICs do not appear. For reference purposes we showed the corresponding
peak labeled as (3) in Figs.\ref{fig:Pic2}(b) and (d). In Fig.\ref{fig:Pic2}(e),
the band structure of the phantom atoms in the presence of BICs is
depicted.

In Fig.\ref{fig:Pic3}(a) we have the enlargement of the region wherein
the peak around $\mathcal{E}\approx-6.3\times10^{-2}D$ is absent
in the LDOS of Fig.\ref{fig:Pic2}(d), thus suggesting the existence
of a BIC at this position. To make this BIC observable, one needs
to introduce the coupling between it and the continuum states, which
can be achieved by the introduction of a small detuning $\triangle\mathcal{E}$
between the energies of the upper and lower adatoms. As a matter of
fact, this detuning will appear automatically due to the hybridizations
of the STM tip with the adatoms, in particular when the former is
found closer to the latter. In Fig.\ref{fig:Pic3}(b) we plot the
LDOS for $\triangle\mathcal{E}=10^{-4}D$. One clearly sees that visible,
although rather weak peak appears at the energy corresponding to the
BIC, in which a true BIC is transformed to a quasi-BIC detectable
in transport experiments. Here we stress that within our theoretical
framework, the role of the detuning $\triangle\mathcal{E}$ is the
emulation of nonperturbative values for the ratio $t_{\text{{d1}}}/t_{\text{{c}}},$
which forces the leaking of the BIC into the system energy continuum
as the aftermath of the renormalization made by the STM tip on the
level of the upper adatom. We should emphasize that Eq.(\ref{eq:LDOSp0})
that describes this energy continuum as well as the conductance $G$
through the system via Eq.(\ref{eq:Conduc}), enclose fingerprints
arising from Eq.(\ref{DOS}) for the adatoms, as for instance, the
quasi-BIC nearby $\mathcal{E}\approx-6.3\times10^{-2}D$ observed
in Fig.\ref{fig:Pic3}(b). This narrow state corresponds to that denoted
by the resonance labeled as $(1)$ in Fig.\ref{fig:Pic2}(b) that
leaks into the continuum of the system.

In the opposite situation where $\triangle\mathcal{E}=0,$ such a
decay of the BIC is prevented due to the mechanism of Fano destructive
interference pointed out previously. Thus Eq.(\ref{eq:LDOSp0}) contains
just the background contribution of Eq.(\ref{eq:phantom_DOS}) at
$\mathcal{E}\approx-6.3\times10^{-2}D$ as Fig.\ref{fig:Pic3}(a)
shows. In this case, the LDOS of the ``graphene+adatoms'' system
behaves as that for the lattice of phantom atoms without adatoms.
Thereby, if sharp resonances appear in both Eqs.(\ref{eq:LDOSp0})
and (\ref{DOS}) at the same position, they reveal the decay of the
state within the adatoms into the energy continuum of the system:
the sharp resonance appearing in the former equation is then considered
a quasi-BIC. A quasi-BIC is characterized by a sharp resonance in
Eq.(\ref{eq:LDOSp0}) being detectable by the conductance $G$ of
Eq.(\ref{eq:Conduc}), which indeed describes at $\mathcal{E}\approx-6.3\times10^{-2}D$
an electron that spends a long time in the vicinity of the adatom,
whose wave function behaves as a Bloch state away from such a site.

Additionally, we clarify that a BIC is represented by the resonance
belonging to the adatom under consideration since it appears via its
DOS given by Eq.(\ref{DOS}), but is absent in Eq.(\ref{eq:LDOSp0})
that determines the system conductance. It is worth noticing that
despite the small but finite width $\Delta\propto|\mathcal{E}|^{3}$
of such a state in Eq.(\ref{DOS}), which is visible around $\mathcal{E}\approx-6.3\times10^{-2}D$
in Fig.\ref{fig:Pic2}(b), the Fano destructive interference mechanism
revealed in this work ensures that the state level is embedded in
the continuum. This situation corresponds to electrons fully trapped
within these adatoms in such a way that the decay rate $\sim\Delta/\hbar$
is suppressed.

On the other hand, the visibility of the quasi-BIC peak can be improved
by approaching both levels $\mathcal{E}_{d}$ and $\mathcal{U}$ towards
the Dirac points $(\mathcal{E}_{d}=-0.02D$ and $\mathcal{U}=0.04D$)
combined with the increasing of the Fermi velocity $v_{F}$ ($v_{F}\approx c/150$)
as shown at panels (c) and (d) of the same figure. From the experimental
perspective the tuning of the Fermi velocity can be performed by changing
the dielectric constant in the substrate hosting the graphene sheet
\cite{TFV,TFV2}. We should point out that the assumption of considering
adatoms slightly off resonance, due to a detuning in energy levels
for detection of quasi-BICs, was adopted in Ref.\,{[}\onlinecite{Gonzalez}{]}
for graphene ribbons. Here we apply the same procedure on our ``graphene+adatoms''
system in order to induce the decay of the BICs within the adatoms
into the continuum of the aforementioned system.

\begin{figure}
\centering{}\includegraphics[width=0.45\textwidth,height=0.45\textheight]{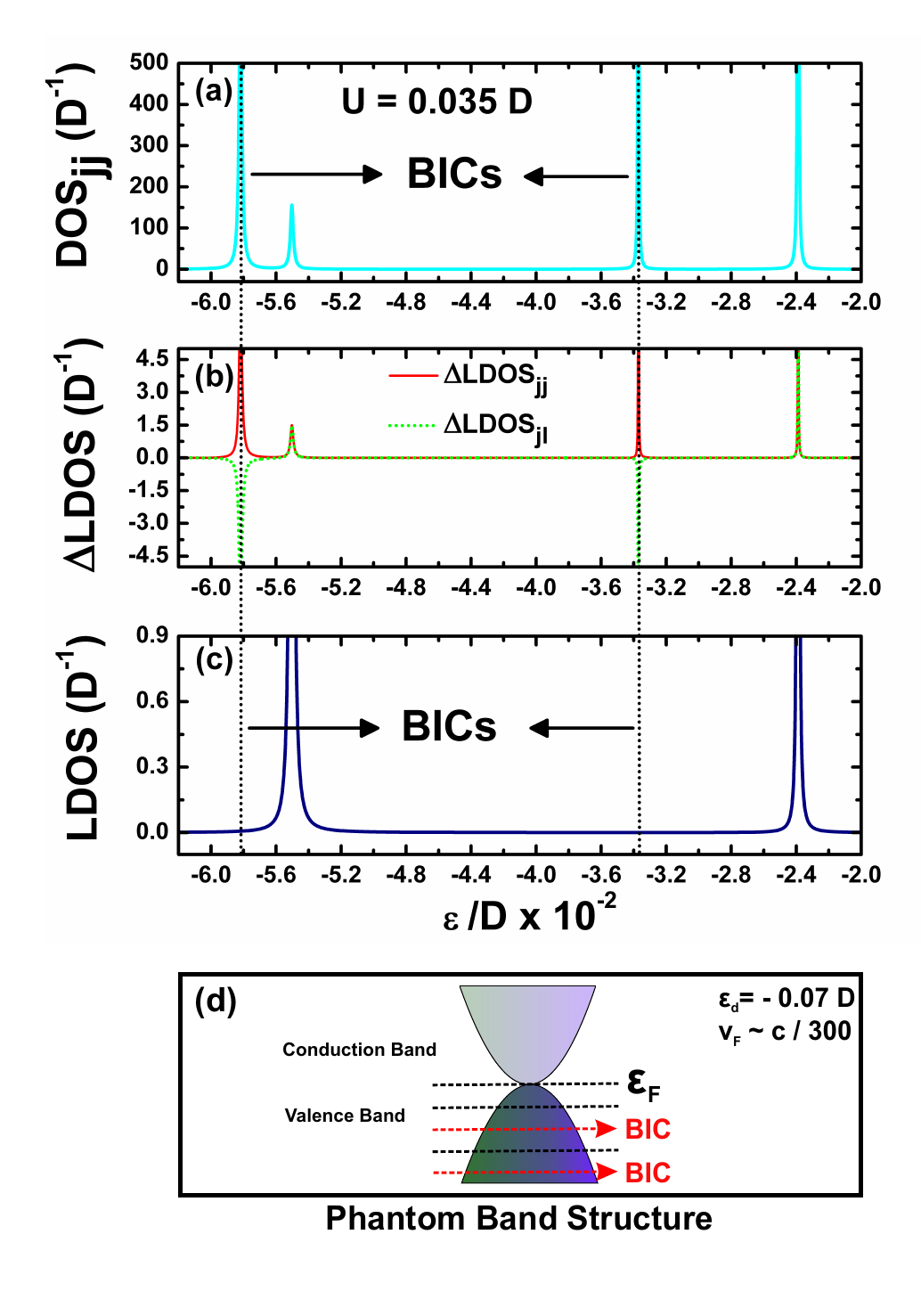}
\protect\protect\protect\protect\protect\protect\protect\protect\protect\protect\protect\protect\protect\caption{\label{fig:Pic4} (Color online) (a) Density of states for the pair
of adatoms $\text{DOS}_{jj}=\text{DOS}_{11}=\text{DOS}_{22}$ within
the valence band. The parameters are $\mathcal{E}_{d}=-0.07D$, $\mathcal{U}=0.035D,$
$\mathcal{V}=0.14D,$ $v_{F}\approx c/300$ and $\triangle\mathcal{E}=0$.
Four peaks are present within the valence band, as $2\mathcal{E}_{d}+\mathcal{U}\protect\neq0$
particle-hole symmetry is broken. (b) Contributions to the LDOS of
graphene from the adatom pair. Diagonal contribution shows four pronounced
peaks, while mixing term shows a couple of pronounced antiresonances.
(c) Total LDOS revealing the BICs (marked by vertical lines) at positions
where resonances of the diagonal term in the LDOS are compensated
by the antiresonances in the mixing term. (d) Sketch of the energy
diagram of the system.}
\end{figure}

In Fig.\ref{fig:Pic4} we present the results for the case of the
broken particle-hole symmetry, taking $\mathcal{U}=0.035D$, $\mathcal{E}_{d}=-0.07D$,
$v_{F}\approx c/300$ and $\triangle\mathcal{E}=0.$ By decreasing
the Coulomb correlation energy from $\mathcal{U}=0.14D$ to $\mathcal{U}=0.035D,$
it is possible to shift the peaks found within the conduction band
$(\mathcal{E}>\mathcal{E}_{F}\equiv0$) for the symmetric case $2\mathcal{E}_{d}+\mathcal{U}=0$
into the valence band $(\mathcal{E}<\mathcal{E}_{F}\equiv0)$ as demonstrated
at the panel (a). Thus instead of the couple of peaks found in Fig.\ref{fig:Pic2}(b),
four resonances appear within the valence band. Although the condition
$2\mathcal{E}_{d}+\mathcal{U}=0$ is not any more satisfied, the underlying
Physics remains: BICs emerge due to Fano antiresonances in the mixing
term of the LDOS that suppress the corresponding peaks found in the
diagonal term (see panels (a), (b) and (c) in which BICs are identified
by vertical lines). Panel (d) shows the band structure in such a case.

\begin{figure}
\centering{}\includegraphics[width=0.45\textwidth,height=0.42\textheight]{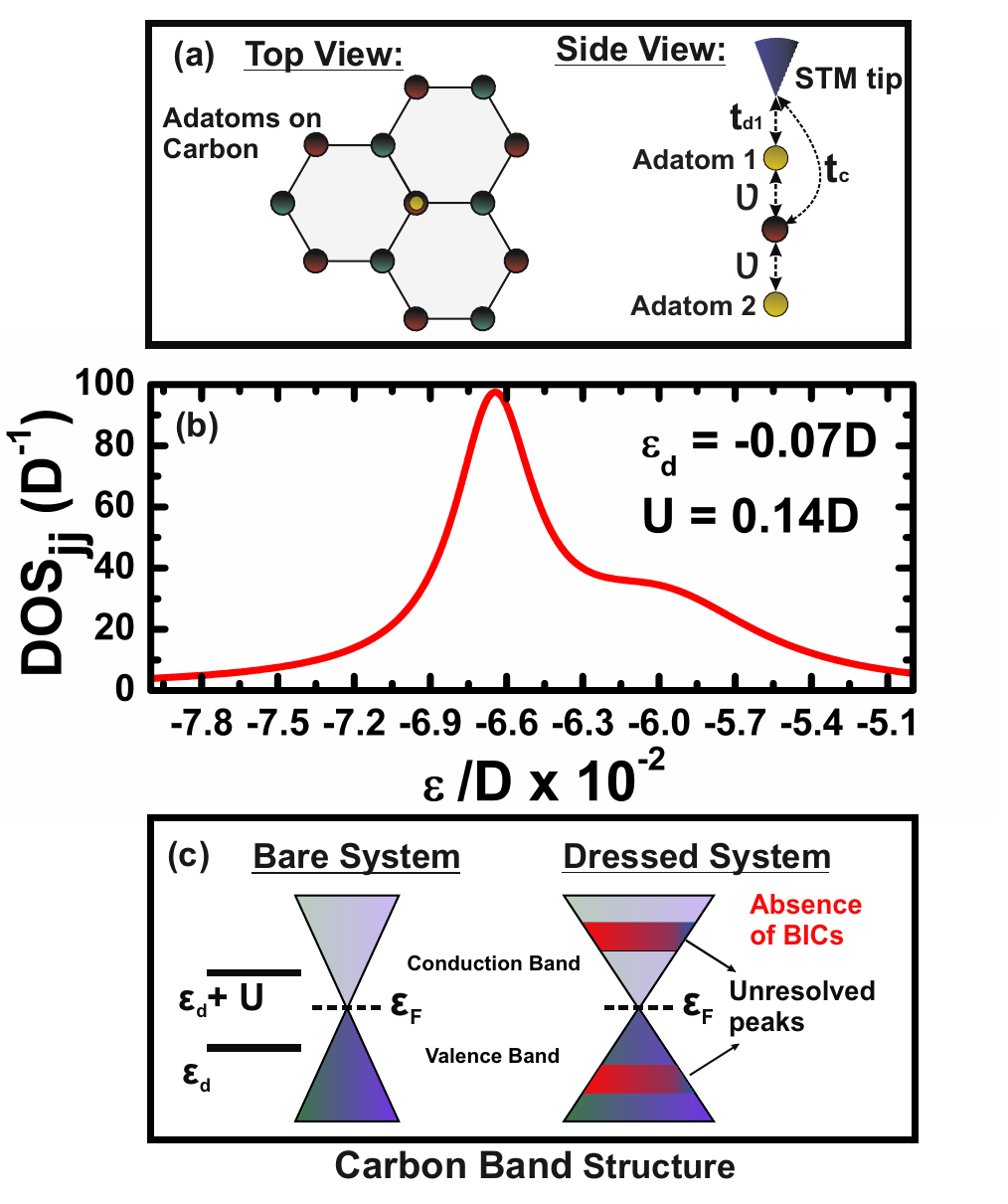}
\protect\protect\protect\protect\protect\protect\protect\protect\protect\protect\protect\protect\protect\caption{\label{fig:Pic5}(Color online) (a) Adatoms aligned with one of the
carbon atoms. (b) $\mathcal{E}_{d}=-0.07D$, $\mathcal{U}=0.14D,$
$\mathcal{V}=0.14D,$ $v_{F}\approx c/300$ and $\triangle\mathcal{E}=0:$
$\text{DOS}_{jj}=\text{DOS}_{11}=\text{DOS}_{22}$ of the adatoms
in which unresolved peaks emerge, thus attesting that BICs can not
be formed in such a geometry. (c) Dirac cones persist exhibiting unresolved
resonances simultaneously within the valence and conduction bands. }
\end{figure}

Fig.\ref{fig:Pic5} depicts the analysis of the situation in which
the adatoms are aligned with one of the carbon atoms of the lattice
as shown in the panel (a). The parameters are the same employed in
Fig.\ref{fig:Pic2}. One can find the expression for the LDOS using
the field operator of a carbon atom \cite{diagonal}
\begin{equation}
\Psi_{\text{{carbon},}\sigma}=\frac{1}{2\pi}\sqrt{\frac{\pi\Omega_{0}}{\mathcal{N}}}\sum_{n}\int\sqrt{\left|k\right|}dkc_{nk\sigma},\label{eq:Carbon}
\end{equation}
instead of the field operator of a phantom atom given by Eq.(\ref{eq:phantom_2}).
The resulting LDOS is given by the same expressions of the Eqs.(\ref{eq:LDOSp0})
and (\ref{eq:LDOSp1}), with only difference that now $\text{\ensuremath{\mathcal{D}}}_{0}\equiv\text{\ensuremath{\mathcal{D}}}_{0}^{\text{{carbon}}}=\Omega_{0}|\mathcal{E}|/2\mathcal{N}\pi(\hbar v_{F})^{2},$\textcolor{blue}{{}
}$\Delta=\pi\text{\ensuremath{\mathcal{D}}}_{0}^{\text{{carbon}}}\mathcal{V}^{2}$
and
\begin{align}
\Sigma_{\text{{carbon}}}({\bold R=\bold0})=\frac{\mathcal{V}^{2}}{D^{2}}\mathcal{E}\ln\Big|\frac{\mathcal{E}^{2}}{D^{2}-\mathcal{E}^{2}}\Big|-i\Delta
\end{align}
stands for the self-energy \cite{Uchoa1} instead of that found in
Eq. (\ref{Sigma0}) for the phantom atom.

Panel (b) shows the DOS for the considered situation. One clearly
sees that differently from the case of the phantom atom BICs do not
appear, since resolved peaks within $\text{DOS}_{jj}=\text{DOS}_{11}=\text{DOS}_{22}$
are absent and a couple of broad merged resonances appears instead.
To explain such a behavior, let us focusing on the Anderson broadening
$\Delta.$ For the case of a phantom atom $\Delta\propto|\mathcal{E}|^{3}$
and as the peaks at $\mathcal{E}\approx-6.3\times10^{-2}D$ and $\mathcal{E}\approx-5\times10^{-2}D$
denoted by (1) and (2) in Fig.\ref{fig:Pic2}(b) are found nearby
the Dirac points $(\mathcal{E}=0),$ they are narrow enough in this
region and can be easily resolved. For the case of the collinear alinement
of the impurities with one of the carbon atoms $\Delta\propto|\mathcal{E}|,$
thereby the broadening of the peaks in the vicinity of the Dirac points
increases and they become unresolved as seen at the Fig.\ref{fig:Pic5}(b).
Fig.\ref{fig:Pic5}(c) displays the sketch of the Dirac cones in such
a situation.

\section{conclusions}

\label{sec:SecIV}

In summary, we have demonstrated that BICs can appear in a system
consisting of a graphene sheet and a collinear pair of adatoms placed
above and below the center of the hexagonal cell, where a fictitious
or phantom atom emulates the six carbon atoms of the cell. The effect
is due to the destructive Fano interference assisted by Coulomb correlations
in the adatoms. We have checked that BICs do not appear if Coulomb
interaction is absent or if adatoms are collinear with one of the
carbon atoms in the lattice.

\section{acknowledgments}

\label{sec:SecV}

This work was supported by the agencies CNPq, CAPES, 2014/14143-0
S{ã}o Paulo Research Foundation (FAPESP), FP7 IRSES project QOCaN
and Rannis project ``Bose and Fermi systems for spintronics''. A.\,C.\,S.
thanks the Nanyang Technological University at Singapore for hospitality.

\end{document}